\documentclass[prl,twocolumn,aps,preprintnumbers,showpacs,superscriptaddress]{revtex4-1}
\usepackage{latexsym}
\usepackage{amsmath}
\usepackage{amssymb}
\usepackage{amsfonts}
\usepackage{ifthen}
\usepackage{revsymb}
\usepackage{yfonts}
\usepackage{graphicx}
\usepackage{graphicx}
\usepackage{amsmath}
\usepackage{amssymb}
\usepackage{amsthm}

\newcommand{\be}{\begin{equation}}
\newcommand{\ee}{\end{equation}}
\newcommand{\ba}{\begin{array}}
\newcommand{\ea}{\end{array}}
\newcommand{\bea}{\begin{eqnarray}}
\newcommand{\eea}{\end{eqnarray}}

\newcommand{\calC}{{\cal C }}

\newcommand{\calL}{{\cal L }}

\newcommand{\calP}{{\cal P }}

\newcommand{\calS}{{\cal S }}

\newcommand{\cG}{{\cal G }}

\newcommand{\calG}{{\cal G }}

\newcommand{\CC}{\mathbb{C}}
\newcommand{\RR}{\mathbb{R}}

\newcommand{\la}{\langle}
\newcommand{\ra}{\rangle}

\newcommand*{\cB}{\mathfrak{B}}
\newcommand*{\supp}{\mathsf{supp}}

\newcommand*{\cP}{\mathcal{P}}

\newtheorem{lemma}{Lemma}

\newtheorem{theorem}{Theorem}

\newtheorem{corol}{Corollary}

\begin{document}

\title{Classification of topologically protected gates
for local stabilizer codes}

\author{Sergey \surname{Bravyi}}
\affiliation{IBM Watson Research
Center,  Yorktown Heights  NY 10598}
\author{Robert \surname{K\"onig}}
\affiliation{IBM Watson Research
Center,  Yorktown Heights  NY 10598}

\date{\today}

\begin{abstract}
Given a quantum error correcting code, an important task is to
find encoded operations that can be implemented efficiently and fault-tolerantly.
In this Letter we focus on topological stabilizer
codes and  encoded unitary gates that can be implemented by a constant-depth quantum circuit.
Such gates have a certain degree of protection
since propagation of errors in a
constant-depth circuit is limited by a constant size light cone.
For the 2D geometry we show that constant-depth
circuits can only implement a finite group of encoded gates
known as the Clifford group.
This implies that  topological protection must be ``turned off"
for at least some steps in the computation in order to achieve universality.
For the 3D geometry we show that an encoded gate $U$ is implementable by a
constant-depth circuit only if $UPU^\dag$ is in the Clifford group  for any
Pauli operator $P$.  This class of gates includes
some non-Clifford gates such as the $\pi/8$ rotation. Our classification applies to
any stabilizer code with geometrically local stabilizers and sufficiently large code distance.
\end{abstract}


\maketitle
Quantum error correcting codes play a central role in all proposed
schemes for fault-tolerant quantum computation. By repeatedly
measuring error syndromes and applying corresponding correction
operations, encoded states can be stored reliably  for
extended periods of time. Furthermore, some codes permit a
fault-tolerant implementation of a computationally universal
set of operations on encoded states~\cite{AGP06}.

Topological codes such as the  surface code
family~\cite{Kitaev03,BK:surface,Dennis01,Fowler08} are arguably
closest to what can currently be achieved in experiments~\cite{divincenzo:JJarchitecture}.
The unifying feature
of all topological codes is the {\em geometric locality} of their check operators.
The physical qubits of a $D$-dimensional topological code
can be laid out on a regular lattice embedded in $\RR^D$ such that
the support of any check operator has diameter $O(1)$.
 The locality ensures that
the syndrome readout requires only short-range quantum
gates and that each qubit participates only in a few gates.
In addition, any topological code has a {\em macroscopic distance}:
a non-trivial operation on encoded states cannot be implemented
by acting on fewer than $d$ qubits, where $d$ can be made arbitrarily large
by increasing the lattice size.
The subclass of
topological stabilizer codes (TSCs) has an additional convenient feature: the
parity check operators are tensor products of single-qubit Pauli
operators. This simple structure allows one to understand properties
of stabilizer codes in much more depth~\cite{NCbook}. The subclass of TSCs
includes the toric and the surface codes~\cite{Kitaev03,BK:surface},
the color codes~\cite{BMD:topo}, as well as
the surface codes with twists~\cite{BombinTwists} or punctured holes~\cite{Fowler08}.
Examples of topological codes which are not TSC are the quantum double models~\cite{Kitaev03} and the Turaev-Viro codes~\cite{Koenig10}.

In this Letter we show that the simple structure of TSCs comes at a price: the set of gates implementable in a fault-tolerant manner is rather restricted for any such code. To
formulate this more precisely, let us say that an encoded gate is {\em
topologically protected} if it can be realized by applying a
constant-depth quantum circuit on
the physical qubits. Here we only consider circuits
with geometrically local gates.  This definition is motivated by the
fact that constant-depth circuits are inherently fault-tolerant: a fault in any
single gate can affect at most $O(1)$ qubits and a pre-existing error
can spread to at most $O(1)$ qubits. Topologically protected gates can
therefore be executed using noisy hardware without introducing too
many errors~\footnote{To prevent errors from accumulating, error correction
needs to be applied after each encoded gate. We implicitly assume that
error correction itself is fault-tolerant for any TSC due to locality of the check operators.}.

To state our main result let us fix the number of logical qubits $k$.
For any $j\ge 1$ define a set of encoded gates
$\calP_j$, $j\ge 1$, such that $\calP_1$ is the group
of $k$-qubit Pauli operators, and $\calP_j$ is a set of
all $k$-qubit unitary operators $U$  such that  $U\calP_1 U^\dag
\subseteq \calP_{j-1}$, where $j\ge 2$. In particular, $\calP_2$ is
the so-called Clifford group, that is,
the group generated by the Hadamard gate $H=(X+Z)/\sqrt{2}$,
the CNOT gate, and the
 $\pi/4$ rotation
$K=\exp{(i\pi Z/4)}$. The set $\calP_3$ includes some
non-Clifford gates such as the $\pi/8$ rotation $\sqrt{K}$. Note that $\calP_j$
is not a group unless $j=1,2$.
Ignoring overall phase factors,
$\calP_j$ is a finite set~\footnote{Any operator $U\in \calP_D$ is specified (up to an overall phase) by a list of $2k$ operators
$\{ UX_jU^\dag, UZ_jU^\dag \in \calP_{D-1}\}$, where $j=1,\ldots,k$.
This shows that $|\calP_D|\le |\calP_{D-1}|^{2k} \le 2^{(2k)^D}$.}.
The sets $\calP_j$  were originally introduced  by Gottesman and
Chuang~\cite{GC99} who proposed a fault-tolerant implementation of
any gate in $\calP_j$ through a recursive application of the gate
teleportation method~\footnote{Our notation is different from Ref.~\cite{GC99}.}.
Surprisingly,  the sets $\calP_j$ also naturally arise in the context
of topological codes. Our main result is the following.

\begin{theorem}
\label{thm:1}
Suppose a unitary operator $U$ implementable by a 
constant-depth  quantum circuit 
preserves the codespace $\calC$ of a topological stabilizer
code on a $D$-dimensional lattice, $D\ge 2$. 
Then the restriction of $U$ onto $\calC$
implements an encoded gate from the set $\calP_D$.
\end{theorem}
Note that the restriction of $U$ onto the codespace $\calC$ can be viewed as a $k$-qubit operator only with respect to some basis of $\calC$. Any stabilizer code has
a basis such that all encoded Pauli operators are products of physical Pauli operators~\cite{NCbook}. We shall always implicitly assume that
such a basis is chosen. Theorem~\ref{thm:1} also holds for any depth-$h$ quantum circuit $U$ with gates of range $r$ such that $\xi,hr \ll d^{1/D}$, where $d$ is the code distance and $\xi$ is the maximum range of the parity check operators. 

An important example of a topologically protected gate is a {\em
transversal gate}: this can be realized by a product of one-qubit
rotations on the physical qubits. For such  a gate, a fault in any
single rotation can affect at most one qubit and pre-existing errors
do not spread to other qubits. A general no-go theorem due to
Eastin and Knill~\cite{EastinKnill2009} asserts that transversal gates
can only generate a finite group  and thus cannot be computationally universal.
In the case of $D$-dimensional TSCs Theorem~\ref{thm:1} provides
a partial characterization of this group by placing it inside $\calP_D$.
Transversality, however, is a rather
restrictive requirement. This motivates the study of  the more general
class of topologically protected gates.  To our knowledge, no
limitations have  previously been derived on the power of
constant-depth quantum circuits in this context.

Let us discuss some implications of Theorem~\ref{thm:1} focusing on the 2D geometry, which is arguably the most practical one. The theorem states
that any topologically protected gate must belong to the Clifford
group $\calP_2$.  Since any quantum circuit composed of Clifford gates can be
efficiently simulated classically~\cite{NCbook},  our result implies that
topological protection  must be ``turned off"
for at least some steps in the computation in order to execute
interesting quantum algorithms. For example, the surface code architecture~\cite{Fowler08,RH:cluster2D} uses injection of  so-called magic states
and distillation techniques to implement non-Clifford gates. The injection step
is not covered by our theorem since the corresponding logical qubit has
no topological protection.  Let us point out that
implementation of non-Clifford gates
is  by far the most time consuming step
in the surface code architecture.
For example, the operational cost of a single $\pi/8$ rotation
 exceeds the one of any topologically protected gate by $2-3$ orders of magnitude~\cite{RHG07}.
Our result suggests that this overhead  cannot be avoided simply by
changing the lattice geometry or using a different  code,
as long as one stays within the class of 2D stabilizer codes.

Our proof of  Theorem~\ref{thm:1} actually covers a more general situation
where one is given two {\em different} codes with codespaces $\calC_1,\calC_2$ and a constant-depth quantum circuit $U$ that maps $\calC_1$ to $\calC_2$. In this case one can view the code $\calC_2$ as a ``local deformation" of the code $\calC_1$.
We prove that $U$ induces an encoded gate from the set
$\calP_D$ provided that both $\calC_1$ and $\calC_2$ are
$D$-dimensional  TSCs.
For the 2D geometry, this shows that any chain of local deformations
$\calC_1\to\calC_2\to \ldots \to \calC_t$ implements an encoded
Clifford group operator
provided that one has uniform bounds on the locality and the distance
of all intermediate codes. Such chains of local deformation can be
used, for instance, to describe braiding of topological defects used
in the surface code architecture to implement encoded CNOT gates~\cite{Fowler08}.

Let us now discuss the case $D\ge 3$. To the best of our knowledge,
the only example of a 3D TSC with topologically protected non-Clifford gates
is the punctured 3D color code due to Bombin and Martin-Delgado~\cite{Bombin06}. It encodes one logical qubit with a transversal  $\pi/8$ rotation which belongs to $\calP_3\backslash \calP_2$.

Theorem~\ref{thm:1} rules
out computational universality of topologically protected gates for $D\ge 3$
in the special case when the number of logical qubits $k$ is a fixed parameter independent of the lattice size $L$. More precisely,
let $\calG_{h,L}$ be the set of encoded gates implementable by
circuits of depth $h$ for a given lattice size $L$ and $\la  \calG_{h,L} \ra$
be the subgroup of the unitary group $U(2^k)$ generated by $\calG_{h,L}$.
By Theorem~\ref{thm:1}, the inclusion $\calG_{h,L} \subseteq \calP_D$
holds for any fixed $h$ and all large enough $L$.
\begin{corol}
\label{cor:1}
Consider any family of $D$-dimensional topological stabilizer codes
defined for an infinite sequence of $L$'s such that
the number of logical qubits $k$ is independent of $L$. Then for any fixed $h$
the inclusion $\la  \calG_{h,L} \ra \subset \calP_D$ holds for all large enough $L$.
\end{corol}
Since the set $\calP_D$ is finite, Corollary~\ref{cor:1} implies that
the set of topologically protected gates cannot be computationally universal
for a family of TSCs with a fixed $k$. Finally, let us point out that restrictions on the transversal $\pi/2^j$ rotation (which belongs to $\calP_j$) similar to the one of Theorem~\ref{thm:1} have been derived for $D$-dimensional color codes by
Bombin et al~\cite{BCHMD09}.

Constant-depth circuits and, more generally, locality preserving
unitary maps play an important role in the classification of different types of topological quantum order in condensed matter physics~\cite{ChenGuWen2010}.
In particular, it was recently shown by
Bombin et al~\cite{Bombin11,Bombin11+}  that any translation-invariant
2D TSC on an infinite lattice is equivalent modulo constant-depth
circuits to one or several copies of the surface code. However, this result does not say anything about topologically protected gates since the latter are only defined in  finite settings. It is also known that constant-depth circuits by themselves are not sufficient
for encoding information into a topological code~\cite{BHV06}.

\begin{figure}[h]
\centerline{\includegraphics[scale=1.0]{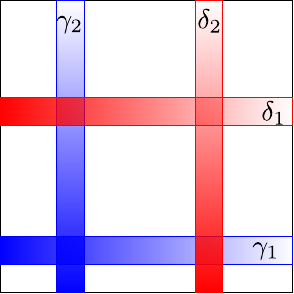}}
\caption{Non-contractible closed strips $\gamma_1,\gamma_2$
and $\delta_1,\delta_2$ on the torus.}
\label{fig:toric}
\end{figure}

In the rest of the paper we prove Theorem~\ref{thm:1} and its corollary.
To illustrate the proof strategy, let us first consider the standard toric
code with two logical qubits.
Recall that logical Pauli operators of the toric code correspond to
non-contractible closed loops on the primal and the dual lattices~\cite{Kitaev03}.
Let~$\gamma_1$ and~$\gamma_2$ be some fixed horizontal and  vertical
non-contractible strips of width~$1$, see Fig.~\ref{fig:toric}. Then
we can choose a complete set of $15$~non-trivial logical Pauli operators supported in~$\gamma\equiv \gamma_1\cup \gamma_2$.
Alternatively, we can choose non-contractible strips
$\delta_1$ and $\delta_2$ as
translations of $\gamma_1$ and $\gamma_2$ respectively
by half the lattice size, see Fig.~\ref{fig:toric}.
Since the toric code is translation-invariant, there exists a
complete set of $15$ logical Pauli operators supported on $\delta\equiv \delta_1\cup \delta_2$.

Consider any unitary operator $U$ implementable by a constant-depth quantum circuit
with short-range gates.
Let $P$ and $Q$ be any pair of logical Pauli operators. We can always find
logical operators $P_\gamma$ and $Q_\delta$
equivalent modulo stabilizers to $P$ and $Q$
such that $P_\gamma$ is supported on~$\gamma$,
while $Q_\delta$ is supported on~$\delta$.
The key observation is that the commutator
\be
\label{K}
K\equiv P_\gamma ( U Q_\delta U^\dag ) P_\gamma^\dag ( U Q_\delta^\dag U^\dag )
\ee
acts non-trivially only on $O(1)$ qubits located near the
intersection of $\gamma$ and $\delta$.  Indeed,
the evolution~$O\mapsto UOU^\dagger$ of any observable $O$
enlarges its support at most by $\rho=hr$, where $h$ is the depth of $U$
and $r$ is the maximum range of its gates.
Loosely speaking, $\rho$ is the radius of a ``light cone'' describing evolution
of observables  under $U$. Note that in our case $\rho=O(1)$.  In particular, $V\equiv U Q_\delta U^\dag$ is supported in $\cB_{\rho}(\delta)$ --- the set of all
qubits within distance $\rho$ from $\delta$. Furthermore, the standard
causality argument implies  that all gates of $U$ lying outside the light cone $\cB_{\rho}(\delta)$ can be omitted without changing $V$.
This shows that $K=P_\gamma V P_\gamma^\dag V^\dag$, where $V$ is a circuit of depth $2h+1$ composed of gates of range $r$. Any gate in $V$
must overlap with the lightcone $\cB_{\rho}(\delta)$.  Here we used the fact that
$Q_\delta$ is a product of single-qubit Pauli operators which can be regarded as a
depth-$1$ circuit. Applying the same causality argument to the evolution
$P_\gamma^\dag \to V P_\gamma^\dag V^\dag$  we conclude that
$K=P_\gamma W P_\gamma^\dag W^\dag$, where $W$ is obtained from
$V$ by omitting all gates lying outside the lightcone of $\gamma$, that is,
$\cB_{r(2h+1)}(\gamma)$. Hence $W$ has support only in $\cB_{O(\rho)}(\gamma \cap \delta)$. The evolution
$W\to P_\gamma W P_\gamma^\dag$ does not enlarge the support of $W$ since
$P_\gamma$ is a product of single-qubit Pauli operators. We conclude that $K$
has support only in $\cB_{O(\rho)}(\gamma \cap \delta)$ which contains only $O(1)$ qubits.

Let $\calC$ be the  four-dimensional codespace
of the toric code and $\Pi$ be the projector onto $\calC$.
By assumption of the theorem, $U$ preserves the codespace $\calC$,
that is, $U\Pi=\Pi U$. Since the operators $U,P_\gamma,Q_\delta$ as well as their Hermitian conjugates preserve $\calC$, we conclude that $K$ preserves
$\calC$ as well.  However, since $K$ acts only on $O(1)$
qubits, the macroscopic distance property implies that $K$
is a trivial logical operator, that is,
\be
K\Pi = c\Pi\label{eq:Kconst}
\ee
for some complex coefficient $c$. We claim that in fact $c=\pm 1$.
Indeed, since $K$ is a unitary operator, one must have $|c|=1$.
Furthermore, Eq.~\eqref{eq:Kconst} can be rewritten
as $VP_\gamma V^\dag \Pi = cP_\gamma \Pi$, where
$V= UQ_\delta U^\dag$. Since $P_\gamma^2=e^{i\theta} I$
for some phase factor $e^{i\theta}$ this implies $e^{i\theta}=c^2e^{i\theta}$,
that is, $c=\pm 1$. To conclude, we have shown that
\be
\label{CR}
P_\gamma (UQ_\delta U^\dag) \Pi = \pm (UQ_\delta U^\dag)P_\gamma \Pi
\ee
for any pair of logical Pauli operators $P,Q$.
Let $\overline{P}$, $\overline{Q}$, and $\overline{U}$
be the encoded two-qubit operators implemented by $P,Q,U$ respectively.
Let $\overline{R}=\overline{U}\cdot \overline{Q} \cdot \overline{U}^\dag$.
From Eq.~(\ref{CR}) one infers that $\overline{P} \cdot \overline{R}=\pm \overline{R} \cdot \overline{P}$. Since  $\overline{P}$ could be an arbitrary two-qubit Pauli operator,
this is possible only if $\overline{R}$ is a Pauli operator itself.
However, since this is true for any Pauli $\overline{Q}$, we conclude that
$\overline{U}$ belongs to the Clifford group.

Let us now consider a more general setting.
We begin by introducing notations and terminology pertaining to
stabilizer codes. Let $n$ be the number of physical qubits
and $\calP(n)$ be group of $n$-qubit Pauli operators.
Any element of $\calP(n)$ has the form
$\gamma P_1\otimes \cdots \otimes P_n$,
where  $P_a\in \{I,X,Y,Z\}$ is a single-qubit Pauli operator
or the identity, and $\gamma\in \CC$ is a phase factor, $|\gamma|=1$.
The set of qubits $a$ on which $P$ acts non-trivially, that is, $P_a\ne I$,
is called the support of $P$ and denoted $\supp(P)$.
A stabilizer code is defined by an abelian {\em stabilizer group}
$\calS\subseteq \calP(n)$ such that $-I\notin \calS$.
Elements of $\calS$ are referred to as {\em stabilizers}.
The corresponding codespace $\calC\subseteq (\CC^2)^{\otimes n}$ is spanned by
 states $\psi$ invariant under the action of any stabilizer,
that is, $S\, \psi=\psi$ for all $S\in \calS$.
We will say that
a stabilizer code has $k$ logical qubits iff $\dim{\calC}=2^k$.
Let $\calL\subset U(2^n)$ be the group of all $n$-qubit unitary operators preserving the codespace $\calC$. Elements of $\calL$ and $\calL\cap \calP(n)$ will be referred to as {\em logical operators} and {\em logical Pauli operators} respectively.

Fixing the basis of the codespace $\calC$
is equivalent to choosing an embedding $J\, : \, (\CC^2)^{\otimes k}
\to (\CC^2)^{\otimes n}$ such that $\calC=\mathrm{Im}(J)$ and
$J^\dag J=I$, that is, $J$ is an isometry.
Note that $\Pi\equiv JJ^\dag$ is the projector onto the codespace.
Given a logical operator
$O\in \calL$, let $\overline{O}\equiv J^\dag O J$ be the
$k$-qubit encoded operator implemented by $O$.
Recall that $\calP_1\equiv \calP(k)$ stands for the group of
$k$-qubit Pauli operators and
\[
{\calP}_j =\{ U\in U(2^k) \, : \, U {\calP}_1 U^\dag \subseteq {\calP}_{j-1}\}
\]
for any $j\ge 2$.  It is well known that for
any stabilizer code one can choose a basis of $\calC$ such that
\[
\calP_1=\{\overline{P}\, : \, P\in \calL\cap \calP(n)\}.
\]
In particular, any encoded Pauli operator
can be implemented by a Pauli operator on the physical qubits.
 A stabilizer code has distance $d$ iff for any logical Pauli operator $P$ supported  on less than $d$ qubits
the encoded operator $\overline{P}$ is proportional to the identity.

Let $\Lambda=[1,L]^D$ be the regular $D$-dimensional cubic lattice
of linear size $L$. Physical qubits occupy sites of $\Lambda$, that is,
$n=L^D$. We will assume that the stabilizer group $\calS$
has a set of local generators $S_1,\ldots, S_{n-k}$ such that
the support of any generator has diameter $\xi=O(1)$,
while the distance of the code $d$ can be made arbitrarily large
by choosing large enough $L$. As it is the case with the toric code,
here we implicitly consider an infinite family of codes defined for a diverging
sequence of $L$'s. A family of stabilizer codes as above will be
called a {\em topological stabilizer code} (TSC).
We will say that a subset of physical qubits $M$ is {\em correctable} iff
for any logical Pauli operator $P$ supported inside $M$ the encoded
operator $\overline{P}$ is proportional to the identity.
By definition, any subset~$M$ of size smaller than the code distance $d$  is correctable. We will use the following facts.
\begin{lemma}[\bf Cleaning Lemma~\cite{BT08}]
Suppose $M$ is a correctable subset of qubits.
Then for any logical Pauli operator $P$
there exists a stabilizer $S$ such that
$PS$ acts trivially on $M$.
\end{lemma}
\begin{lemma}[\bf Union Lemma~\cite{BPT10,HP10}]
Suppose~$M$ and~$K$ are disjoint correctable subsets of qubits
such that the distance between~$M$ and~$K$ is greater than
the diameter~$\xi$ of the stabilizer generators.
Then the union $M\cup K$ is correctable.
\end{lemma}

Suppose $\calS_1$ and $\calS_2$ are TSCs defined on the same lattice $\Lambda$
and encoding the same number of logical qubits.
Let $\Pi_1=J_1J_1^\dag$ and $\Pi_2=J_2J_2^\dag$ be the projectors
onto the codespaces of $\calS_1$
and $\calS_2$ respectively.  We will say that
a unitary operator $U\in U(2^n)$ is a
{\em morphism} between $\calS_1$ and $\calS_2$
iff $U$ maps the codespace of $\calS_1$ to the codespace of $\calS_2$, that is,
\be
\label{mor}
U\Pi_1U^\dag = \Pi_2.
\ee
We will say that $U$ is a $\calP_j$-morphism iff it
implements an encoded element of~${\calP}_j$ on the respective codespaces,
that is,
\be
\label{Cmor}
\hat{U}\equiv J_2^\dag U J_1 \in {\calP}_j.
\ee
Theorem~\ref{thm:1}, in its more general form, states the following:
 {\em if~$U$ is a morphism between $D$-dimensional TSCs $\calS_1$ and~$\calS_2$, and $U$ is implementable by a constant-depth quantum circuit with short-range gates, then~$U$ is a~$\calP_{D}$-morphism for all large enough~$L$.}

We proceed to the proof for $D=2$
in which case we need to show that $\hat{U}$ is in the Clifford group.
Let $Q$ and $P$ be any logical Pauli operators for the codes
$\calS_1$ and $\calS_2$ respectively. Let
\[
\overline{Q}=J_1^\dag Q J_1 \quad \mbox{and} \quad \overline{P}=J_2^\dag P J_2.
\]
Recall that $\overline{Q}$
and $\overline{P}$ could be any $k$-qubit Pauli operators.
As in the case of the toric code, we will examine a commutator
\[
K=P (UQU^\dag) P^\dag (U Q^\dag U^\dag)
\]
and prove that the restriction of $K$ onto the codespace of $\calS_2$
is proportional to the identity, namely,
\be
\label{goal1}
K\Pi_2 = \pm \Pi_2.
\ee
This can be rewritten as $J_2^\dag K J_2=\pm I$.
Using the identities $J_2^\dag P=\overline{P} J_2^\dag$,
$J_1^\dag Q=\overline{Q} J_1^\dag$, and $J_2^\dag U=\hat{U} J_1^\dag$
one easily gets
\[
J_2^\dag K J_2=\overline{P} (\hat{U} \overline{Q} \hat{U}^\dag) \overline{P}^\dag
(\hat{U} \overline{Q}^\dag \hat{U}^\dag)=\pm I.
\]
This shows that $\hat{U} \overline{Q} \hat{U}^\dag$
either commutes or anti-commutes with any Pauli operator.
This is possible only if $\hat{U} \overline{Q} \hat{U}^\dag$
is a Pauli operator itself. Since this holds for any Pauli $\overline{Q}$,
we conclude that $\hat{U}\in \calP_2$, that is, $\hat{U}$ is in the Clifford group.

\begin{figure}[h]
\centerline{\includegraphics[scale=0.6]{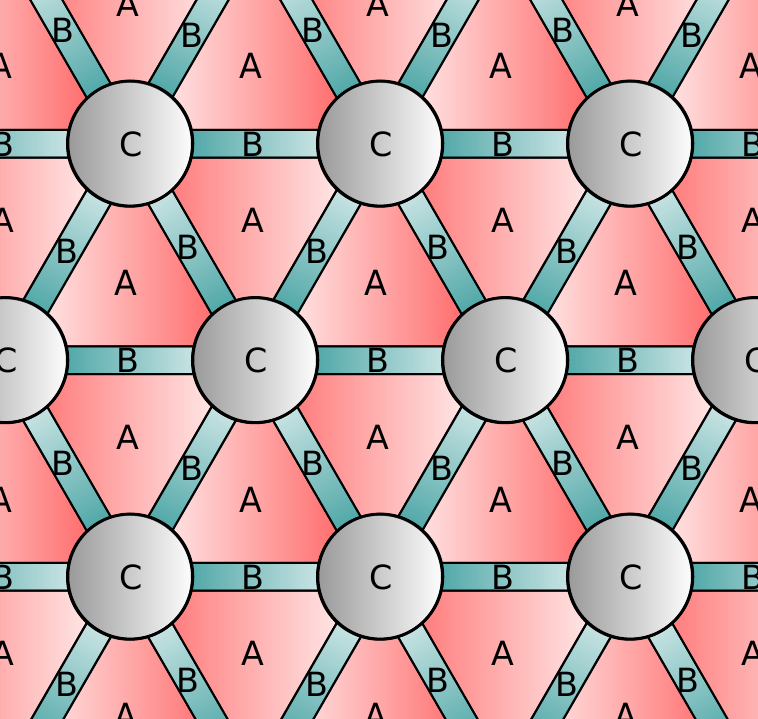}}
\caption{Simplicial partition of the lattice $\Lambda=ABC$. Starting from a triangulation with regular triangles having sides of length~$R$, let $C$ be the union
of discs of radius~$R/4$ centered on the vertices of the triangulation.
 Let $B\subset \Lambda\backslash C$  be union of the $R/8$-neighborhoods of each edge in the remaining surface. Finally, let $A=\Lambda\backslash (B\cup C)$ be the union of the remaining capped triangles.  A similar but rectangular partition has been used in~\cite{BPT10} to derive upper bounds on parameters of TSCs, but is less suitable for generalization to~$D>2$.  }
\label{fig:abc}
\end{figure}

It remains to prove Eq.~(\ref{goal1}).
For any integer $1\ll R\ll L$  the lattice can be partitioned into three disjoint regions, $\Lambda=ABC$,
such that each region $A=\cup_i A_i,B=\cup_j B_j,C=\cup_k C_k$ consists of disjoint chunks of diameter~$O(R)$ separated by distance~$\Omega(R)$, see Fig.~\ref{fig:abc} for an example. We assume that the lattice is large enough so we can choose
$\xi,hr\ll R\ll \sqrt{d}$ (recall that $r$~denotes the range of the gates in $U$, whereas
$h$ is the depth of $U$).

This choice guarantees for any $\rho=O(hr)$, the $\rho$-neighborhood $\cB_{\rho}(A_j)$ of any chunk $A_j$ contains fewer qubits than the code distance~$d$, hence~$\cB_{\rho}(A_j)$ is correctable. Furthermore, since the separation between
$\cB_{\rho}(A_i)$ and $\cB_{\rho}(A_j)$
with $i\ne j$ is larger than $\xi$, the Union Lemma implies that the entire region~$\cB_{\rho}(A)=\cup_i \cB_{\rho}(A_i)$ is correctable.  In a similar fashion, we can show that the regions $\cB_{\rho}(B)$ and $C$ are correctable.

Applying the Cleaning Lemma to the logical Pauli operator
$Q$ and the region $\cB_\rho(A)$ we can find a stabilizer $S_1\in \calS_1$ such that $QS_1$ acts trivially on $\cB_\rho(A)$.
Applying the same arguments to the logical Pauli operator $P$ and the region $B$ we can find a stabilizer $S_2\in \calS_2$ such that $PS_2$
acts trivially on $\cB_\rho(B)$. Replacing $Q$ and $P$ by equivalent logical operator $QS_1$ and $PS_2$ (which does not change $\overline{Q}$ and $\overline{P}$)
we can now assume that
\[
\supp(Q)\cap \cB_\rho(A) =\emptyset   \quad \mbox{and} \quad
\supp(P) \cap \cB_\rho(B)=\emptyset.
\]
Consider the evolution $Q\to U Q U^\dag$. It
enlarges the support of $Q$ at most by $rh<\rho$,
so that the light cone of $Q$ and all gates of $U$
overlapping with it are contained in $BC$.
Applying the causality argument used in the toric code example
we conclude that $U Q U^\dag$
can be implemented by a depth-$(2h+1)$ circuit $V$ with gates of range $r$
and all gates of $V$ are supported in $BC$.
Note that $K=PVP^\dag V^\dag$.
Applying the causality argument to the time evolution
$P^\dag \to VP^\dag V^\dag$ which is characterized by
a light cone of radius $r(2h+1)<\rho$, we conclude that
$K=PWP^\dag W^\dag$, where $W$ is obtained from $V$
by omitting all gates lying outside the light cone of $\supp{(P)}$.
Our assumptions on $\supp{(P)}$ imply that any gate
supported in $B$ or overlapping with $B$ lies outside
the light cone of $\supp{(P)}$. Hence all gates of $W$
are supported in $C$.
The evolution $W\to  PWP^\dag$ does not enlarge the support of $W$
since $P$ is a product of single-qubit Pauli operators.
We conclude that $K$ is supported in $C$
which is a correctable region as argued above.
Using Eq.~(\ref{mor}) one can easily check that $K$
preserves the codespace of $\calS_2$, that is,
$K \Pi_2 = \Pi_2 K$.
Let $K=\sum_\alpha c_\alpha K_\alpha$ be the
expansion of $K$ in the basis of Pauli operators,
where $c_\alpha$ are complex coefficients, and $K_\alpha$
are $n$-qubit Pauli operators. Note that all $K_\alpha$ are supported in $C$.
Then
\[
K\Pi_2 = \Pi_2 K \Pi_2 =\sum_\alpha c_\alpha \Pi_2 K_\alpha \Pi_2.
\]
Since $C$ is a correctable region, $\Pi_2 K_\alpha \Pi_2 = x_\alpha \Pi_2$ for some complex coefficient~$x_\alpha$.
This shows that $K\Pi_2=c\Pi_2$ for some coefficient $c$.
The same arguments as in the toric code example show that $c=\pm 1$. This proves Eq.~(\ref{goal1}) and completes the proof of the theorem for $D=2$.

Let us briefly sketch the generalization to $D>2$. As before, we can partition the lattice $\Lambda=\cup_{j=1}^{D+1} \Lambda_j$ into $D+1$~regions such that each region $\Lambda_j$
is a disjoint union of chunks of size $O(R)$ separated by distance $\Omega(R)$
for some $\xi,hr\ll R\ll d^{1/D}$.
Each region~$\Lambda_j$ is correctable by the Union Lemma.
The desired partition can be constructed by analogy with the
2D case, see Fig.~\ref{fig:abc}, such that $\Lambda_j$
corresponds to $(j-1)$-dimensional simplices in a triangulation of $\Lambda$.
Let $P_1$ be any logical Pauli operator for the code $\calS_1$ and
$P_2,\ldots,P_D$ be any logical Pauli operator for the code $\calS_2$.
By applying the Cleaning Lemma for all $j=1,\ldots,D$, we can assume that
$\supp{(P_j)}$ does not overlap with $\rho$-neighborhood of $\Lambda_j$
for some $\rho=O(hr)$.  Define $K_1=UP_1U^\dagger$ and $K_j=P_j^\dagger K_{j-1}P_j K_{j-1}^\dagger$ for $j=2,\ldots, D$. Starting from $K_1$, we can proceed inductively to argue (using the causality argument) that $K_j$ acts trivially on $\cup_{k=1}^j\Lambda_k$. In particular, $K_D$ is supported on the correctable region $\Lambda_{D+1}$, and we conclude that $\overline{K}_D=\pm I$ for any choice of logical Pauli operators $\{P_j\}_j$. This implies that $\overline{K}_{D-1}\in{\cP}_1$ is a Pauli operator. We then proceed inductively:  suppose we have shown that
$\overline{K}_{D-j}\in{\cP}_j$. The fact that $\overline{P}_{D-j}\overline{K}_{D-j}\in {\cP}_1{\cP}_{j}
\subset {\cP}_j$ and the definition of $K_{D-j}$ then imply that $\overline{K}_{D-j-1}\overline{P}_{D-j}\overline{K}_{D-j-1}^\dagger\in{\cP}_j $ and hence $\overline{K}_{D-j-1}\in{\cP}_{j+1}$. In particular, we have
$\overline{K}_1\in {\cP}_{D-1}$ and hence $\hat{U}\in{\cP}_D$ as claimed.

We conclude by  proving Corollary~\ref{cor:1}.
Consider any subset $\cG\subset\cP_D$
and let $\la \cG \ra\subseteq U(2^k)$ be the group generated by $\cG$.
Suppose $\langle\cG\rangle$ is not contained in $\cP_D$.
Let $s=s(\cG)$ be the smallest integer such that $U_1\cdots U_s\notin \cP_D$
for some $U_1,\ldots,U_s\in \cG$. If $\langle\cG\rangle\subseteq \cP_D$,
define $s(\cG)=0$. Define
\begin{align*}
s_*=\max_{\cG\subseteq\cP_D} s(\cG)\ .
\end{align*}
Because~$\cP_D$ is a finite set (ignoring overall phase factors)
and depends only on the number of
logical qubits $k$, we conclude that  $s_*=s_*(k)$ is well-defined.

Suppose  that the set of protected gates implementable by a depth-$h$
circuit generates unitaries not belonging to~$\cP_D$, i.e.,~$\langle
\cG_{h,L}\rangle\not\subset\cP_D$. By definition of $s_*$ and because
$\cG_{h,L}\subset\cP_D$ by Theorem~\ref{thm:1},   there is an element
$U\in\langle\cG_{h,L}\rangle$, $U\not\in\cP_D$ such that $U=U_1\cdots
U_{s'}$ can be written as a product of $s'\leq s_*$
factors~$U_j\in\cG_{h,L}$. We conclude that~$U\in\cG_{s_*\cdot h,L}$
is an encoded gate implementable by a depth-$s_*\cdot h=O(1)$ circuit,
hence~$U\in\cP_D$ by Theorem~\ref{thm:1}, which is a contradiction.

\section{Acknowledgments}
This work was partially supported by the
DARPA QUEST program under contract number HR0011-09-C-0047
and by the Intelligence Advanced Research Projects Activity (IARPA) via Department of Interior National Business Center contract number D11PC20167. The U.S. Government is authorized to reproduce and distribute reprints for Governmental purposes notwithstanding any copyright annotation thereon. Disclaimer: The views and conclusions contained herein are those of the authors and should not be interpreted as necessarily representing the official policies or endorsements, either expressed or implied, of IARPA, DoI/NBC, or the U.S. Government.


\begin{thebibliography}{26}%
\makeatletter
\providecommand \@ifxundefined [1]{%
 \@ifx{#1\undefined}
}%
\providecommand \@ifnum [1]{%
 \ifnum #1\expandafter \@firstoftwo
 \else \expandafter \@secondoftwo
 \fi
}%
\providecommand \@ifx [1]{%
 \ifx #1\expandafter \@firstoftwo
 \else \expandafter \@secondoftwo
 \fi
}%
\providecommand \natexlab [1]{#1}%
\providecommand \enquote  [1]{``#1''}%
\providecommand \bibnamefont  [1]{#1}%
\providecommand \bibfnamefont [1]{#1}%
\providecommand \citenamefont [1]{#1}%
\providecommand \href@noop [0]{\@secondoftwo}%
\providecommand \href [0]{\begingroup \@sanitize@url \@href}%
\providecommand \@href[1]{\@@startlink{#1}\@@href}%
\providecommand \@@href[1]{\endgroup#1\@@endlink}%
\providecommand \@sanitize@url [0]{\catcode `\\12\catcode `\$12\catcode
  `\&12\catcode `\#12\catcode `\^12\catcode `\_12\catcode `\%12\relax}%
\providecommand \@@startlink[1]{}%
\providecommand \@@endlink[0]{}%
\providecommand \url  [0]{\begingroup\@sanitize@url \@url }%
\providecommand \@url [1]{\endgroup\@href {#1}{\urlprefix }}%
\providecommand \urlprefix  [0]{URL }%
\providecommand \Eprint [0]{\href }%
\@ifxundefined \urlstyle {%
  \providecommand \doi  [0]{\begingroup \@sanitize@url \@doi}%
  \providecommand \@doi [1]{\endgroup \@@startlink {\doibase
  #1}doi:\discretionary {}{}{}#1\@@endlink }%
}{%
  \providecommand \doi  [0]{doi:\discretionary{}{}{}\begingroup
  \urlstyle{rm}\Url }%
}%
\providecommand \doibase [0]{http://dx.doi.org/}%
\providecommand \Doi [0]{\begingroup \@sanitize@url \@Doi }%
\providecommand \@Doi  [1]{\endgroup\@@startlink{\doibase#1}\@@Doi}%
\providecommand \@@Doi [1]{#1\@@endlink}%
\providecommand \selectlanguage [0]{\@gobble}%
\providecommand \bibinfo  [0]{\@secondoftwo}%
\providecommand \bibfield  [0]{\@secondoftwo}%
\providecommand \translation [1]{[#1]}%
\providecommand \BibitemOpen [0]{}%
\providecommand \bibitemStop [0]{}%
\providecommand \bibitemNoStop [0]{.\EOS\space}%
\providecommand \EOS [0]{\spacefactor3000\relax}%
\providecommand \BibitemShut  [1]{\csname bibitem#1\endcsname}%
\bibitem [{\citenamefont {Aliferis}\ \emph {et~al.}(2006)\citenamefont
  {Aliferis}, \citenamefont {Gottesman},\ and\ \citenamefont
  {Preskill}}]{AGP06}%
  \BibitemOpen
  \bibfield  {author} {\bibinfo {author} {\bibfnamefont {P.}~\bibnamefont
  {Aliferis}}, \bibinfo {author} {\bibfnamefont {D.}~\bibnamefont {Gottesman}},
  \ and\ \bibinfo {author} {\bibfnamefont {J.}~\bibnamefont {Preskill}},\
  }\href@noop {} {\bibfield  {journal} {\bibinfo  {journal} {Quant. Inf.
  Comput.},\ }\textbf {\bibinfo {volume} {6}},\ \bibinfo {pages} {97} (\bibinfo
  {year} {2006})}\BibitemShut {NoStop}%
\bibitem [{\citenamefont {Kitaev}(2003)}]{Kitaev03}%
  \BibitemOpen
  \bibfield  {author} {\bibinfo {author} {\bibfnamefont {A.~Y.}\ \bibnamefont
  {Kitaev}},\ }\href@noop {} {\bibfield  {journal} {\bibinfo  {journal} {Annals
  of Physics},\ }\textbf {\bibinfo {volume} {303}},\ \bibinfo {pages} {2}
  (\bibinfo {year} {2003})}\BibitemShut {NoStop}%
\bibitem [{\citenamefont {{Bravyi}}\ and\ \citenamefont
  {{Kitaev}}(1998)}]{BK:surface}%
  \BibitemOpen
  \bibfield  {author} {\bibinfo {author} {\bibfnamefont {S.}~\bibnamefont
  {{Bravyi}}}\ and\ \bibinfo {author} {\bibfnamefont {A.~Y.}\ \bibnamefont
  {{Kitaev}}},\ }\href@noop {} {\bibfield  {journal} {\bibinfo  {journal}
  {ArXiv quant-ph/9811052}} (\bibinfo {year} {1998})}\BibitemShut {NoStop}%
\bibitem [{\citenamefont {Dennis}\ \emph {et~al.}(2002)\citenamefont {Dennis},
  \citenamefont {Kitaev}, \citenamefont {Landahl},\ and\ \citenamefont
  {Preskill}}]{Dennis01}%
  \BibitemOpen
  \bibfield  {author} {\bibinfo {author} {\bibfnamefont {E.}~\bibnamefont
  {Dennis}}, \bibinfo {author} {\bibfnamefont {A.}~\bibnamefont {Kitaev}},
  \bibinfo {author} {\bibfnamefont {A.}~\bibnamefont {Landahl}}, \ and\
  \bibinfo {author} {\bibfnamefont {J.}~\bibnamefont {Preskill}},\ }\href@noop
  {} {\bibfield  {journal} {\bibinfo  {journal} {J. Math. Phys.},\ }\textbf
  {\bibinfo {volume} {43}},\ \bibinfo {pages} {4452} (\bibinfo {year}
  {2002})}\BibitemShut {NoStop}%
\bibitem [{\citenamefont {Fowler}\ \emph {et~al.}(2009)\citenamefont {Fowler},
  \citenamefont {Stephens},\ and\ \citenamefont {Groszkowski}}]{Fowler08}%
  \BibitemOpen
  \bibfield  {author} {\bibinfo {author} {\bibfnamefont {A.~G.}\ \bibnamefont
  {Fowler}}, \bibinfo {author} {\bibfnamefont {A.~M.}\ \bibnamefont
  {Stephens}}, \ and\ \bibinfo {author} {\bibfnamefont {P.}~\bibnamefont
  {Groszkowski}},\ }\href@noop {} {\bibfield  {journal} {\bibinfo  {journal}
  {Phys. Rev. A},\ }\textbf {\bibinfo {volume} {80}},\ \bibinfo {pages}
  {052312} (\bibinfo {year} {2009})}\BibitemShut {NoStop}%
\bibitem [{\citenamefont {{DiVincenzo}}(2009)}]{divincenzo:JJarchitecture}%
  \BibitemOpen
  \bibfield  {author} {\bibinfo {author} {\bibfnamefont {D.~P.}\ \bibnamefont
  {{DiVincenzo}}},\ }\href@noop {} {\bibfield  {journal} {\bibinfo  {journal}
  {Physica Scripta Volume T},\ }\textbf {\bibinfo {volume} {137}},\ \bibinfo
  {pages} {014020} (\bibinfo {year} {2009})}\BibitemShut {NoStop}%
\bibitem [{\citenamefont {Nielsen}\ and\ \citenamefont
  {Chuang}(2000)}]{NCbook}%
  \BibitemOpen
  \bibfield  {author} {\bibinfo {author} {\bibfnamefont {M.~A.}\ \bibnamefont
  {Nielsen}}\ and\ \bibinfo {author} {\bibfnamefont {I.~L.}\ \bibnamefont
  {Chuang}},\ }\href@noop {} {\emph {\bibinfo {title} {Quantum Computation and
  Quantum Information}}}\ (\bibinfo  {publisher} {Cambridge University Press},\
  \bibinfo {year} {2000})\BibitemShut {NoStop}%
\bibitem [{\citenamefont {Bombin}\ and\ \citenamefont
  {Martin-Delgado}(2006)}]{BMD:topo}%
  \BibitemOpen
  \bibfield  {author} {\bibinfo {author} {\bibfnamefont {H.}~\bibnamefont
  {Bombin}}\ and\ \bibinfo {author} {\bibfnamefont {M.~A.}\ \bibnamefont
  {Martin-Delgado}},\ }\href@noop {} {\bibfield  {journal} {\bibinfo  {journal}
  {Phys. Rev. Lett.},\ }\textbf {\bibinfo {volume} {97}} (\bibinfo {year}
  {2006})}\BibitemShut {NoStop}%
\bibitem [{\citenamefont {Bombin}(2010)}]{BombinTwists}%
  \BibitemOpen
  \bibfield  {author} {\bibinfo {author} {\bibfnamefont {H.}~\bibnamefont
  {Bombin}},\ }\href@noop {} {\bibfield  {journal} {\bibinfo  {journal} {Phys.
  Rev. Lett.},\ }\textbf {\bibinfo {volume} {105}},\ \bibinfo {pages} {030403}
  (\bibinfo {year} {2010})}\BibitemShut {NoStop}%
\bibitem [{\citenamefont {Koenig}\ \emph {et~al.}(2010)\citenamefont {Koenig},
  \citenamefont {Kuperberg},\ and\ \citenamefont {Reichardt}}]{Koenig10}%
  \BibitemOpen
  \bibfield  {author} {\bibinfo {author} {\bibfnamefont {R.}~\bibnamefont
  {Koenig}}, \bibinfo {author} {\bibfnamefont {G.}~\bibnamefont {Kuperberg}}, \
  and\ \bibinfo {author} {\bibfnamefont {B.~W.}\ \bibnamefont {Reichardt}},\
  }\href@noop {} {\bibfield  {journal} {\bibinfo  {journal} {Ann. of Phys.},\
  }\textbf {\bibinfo {volume} {325}},\ \bibinfo {pages} {2707} (\bibinfo {year}
  {2010})}\BibitemShut {NoStop}%
\bibitem [{Note1()}]{Note1}%
  \BibitemOpen
  \bibinfo {note} {To prevent errors from accumulating, error correction needs
  to be applied after each encoded gate. We implicitly assume that error
  correction itself is fault-tolerant for any TSC due to locality of the check
  operators.}\BibitemShut {Stop}%
\bibitem [{Note2()}]{Note2}%
  \BibitemOpen
  \bibinfo {note} {Any operator $U\in {\protect \cal P }_D$ is specified (up to
  an overall phase) by a list of $2k$ operators $\protect \{ UX_jU^\protect
  \dag , UZ_jU^\protect \dag \in {\protect \cal P }_{D-1}\protect \}$, where
  $j=1,\protect \ldots ,k$. This shows that $|{\protect \cal P }_D|\le
  |{\protect \cal P }_{D-1}|^{2k} \le 2^{(2k)^D}$.}\BibitemShut {Stop}%
\bibitem [{\citenamefont {Gottesman}\ and\ \citenamefont
  {Chuang}(1999)}]{GC99}%
  \BibitemOpen
  \bibfield  {author} {\bibinfo {author} {\bibfnamefont {D.}~\bibnamefont
  {Gottesman}}\ and\ \bibinfo {author} {\bibfnamefont {I.~L.}\ \bibnamefont
  {Chuang}},\ }\href@noop {} {\bibfield  {journal} {\bibinfo  {journal}
  {Nature},\ }\textbf {\bibinfo {volume} {402}},\ \bibinfo {pages} {390}
  (\bibinfo {year} {1999})}\BibitemShut {NoStop}%
\bibitem [{Note3()}]{Note3}%
  \BibitemOpen
  \bibinfo {note} {Our notation is different from Ref.~\cite
  {GC99}.}\BibitemShut {Stop}%
\bibitem [{\citenamefont {Eastin}\ and\ \citenamefont
  {Knill}(2009)}]{EastinKnill2009}%
  \BibitemOpen
  \bibfield  {author} {\bibinfo {author} {\bibfnamefont {B.}~\bibnamefont
  {Eastin}}\ and\ \bibinfo {author} {\bibfnamefont {E.}~\bibnamefont {Knill}},\
  }\href@noop {} {\bibfield  {journal} {\bibinfo  {journal} {Phys. Rev.
  Lett.},\ }\textbf {\bibinfo {volume} {102}},\ \bibinfo {pages} {110502}
  (\bibinfo {year} {2009})}\BibitemShut {NoStop}%
\bibitem [{\citenamefont {{Raussendorf}}\ and\ \citenamefont
  {{Harrington}}(2007)}]{RH:cluster2D}%
  \BibitemOpen
  \bibfield  {author} {\bibinfo {author} {\bibfnamefont {R.}~\bibnamefont
  {{Raussendorf}}}\ and\ \bibinfo {author} {\bibfnamefont {J.}~\bibnamefont
  {{Harrington}}},\ }\href@noop {} {\bibfield  {journal} {\bibinfo  {journal}
  {Phys. Rev. Lett.},\ }\textbf {\bibinfo {volume} {98}},\ \bibinfo {pages}
  {190504} (\bibinfo {year} {2007})}\BibitemShut {NoStop}%
\bibitem [{\citenamefont {{Raussendorf}}\ \emph {et~al.}(2007)\citenamefont
  {{Raussendorf}}, \citenamefont {{Harrington}},\ and\ \citenamefont
  {{Goyal}}}]{RHG07}%
  \BibitemOpen
  \bibfield  {author} {\bibinfo {author} {\bibfnamefont {R.}~\bibnamefont
  {{Raussendorf}}}, \bibinfo {author} {\bibfnamefont {J.}~\bibnamefont
  {{Harrington}}}, \ and\ \bibinfo {author} {\bibfnamefont {K.}~\bibnamefont
  {{Goyal}}},\ }\href@noop {} {\bibfield  {journal} {\bibinfo  {journal} {New
  J. Phys.},\ }\textbf {\bibinfo {volume} {9}},\ \bibinfo {pages} {199}
  (\bibinfo {year} {2007})}\BibitemShut {NoStop}%
\bibitem [{\citenamefont {Bombin}\ and\ \citenamefont
  {Martin-Delgado}(2007)}]{Bombin06}%
  \BibitemOpen
  \bibfield  {author} {\bibinfo {author} {\bibfnamefont {H.}~\bibnamefont
  {Bombin}}\ and\ \bibinfo {author} {\bibfnamefont {M.}~\bibnamefont
  {Martin-Delgado}},\ }\href@noop {} {\bibfield  {journal} {\bibinfo  {journal}
  {Phys.Rev.Lett.},\ }\textbf {\bibinfo {volume} {98}},\ \bibinfo {pages}
  {160502} (\bibinfo {year} {2007})}\BibitemShut {NoStop}%
\bibitem [{\citenamefont {Bombin}\ \emph {et~al.}(2009)\citenamefont {Bombin},
  \citenamefont {Chhajlany}, \citenamefont {Horodecki},\ and\ \citenamefont
  {Martin-Delgado}}]{BCHMD09}%
  \BibitemOpen
  \bibfield  {author} {\bibinfo {author} {\bibfnamefont {H.}~\bibnamefont
  {Bombin}}, \bibinfo {author} {\bibfnamefont {R.~W.}\ \bibnamefont
  {Chhajlany}}, \bibinfo {author} {\bibfnamefont {M.}~\bibnamefont
  {Horodecki}}, \ and\ \bibinfo {author} {\bibfnamefont {M.}~\bibnamefont
  {Martin-Delgado}},\ }\href@noop {} {\bibfield  {journal} {\bibinfo  {journal}
  {arXiv:0907.5228}} (\bibinfo {year} {2009})}\BibitemShut {NoStop}%
\bibitem [{\citenamefont {Chen}\ \emph {et~al.}(2010)\citenamefont {Chen},
  \citenamefont {Gu},\ and\ \citenamefont {Wen}}]{ChenGuWen2010}%
  \BibitemOpen
  \bibfield  {author} {\bibinfo {author} {\bibfnamefont {X.}~\bibnamefont
  {Chen}}, \bibinfo {author} {\bibfnamefont {Z.-C.}\ \bibnamefont {Gu}}, \ and\
  \bibinfo {author} {\bibfnamefont {X.-G.}\ \bibnamefont {Wen}},\ }\href@noop
  {} {\bibfield  {journal} {\bibinfo  {journal} {Phys. Rev. B},\ }\textbf
  {\bibinfo {volume} {82}},\ \bibinfo {pages} {155138} (\bibinfo {year}
  {2010})}\BibitemShut {NoStop}%
\bibitem [{\citenamefont {Bombin}\ \emph {et~al.}(2011)\citenamefont {Bombin},
  \citenamefont {Duclos-Cianci},\ and\ \citenamefont {Poulin}}]{Bombin11}%
  \BibitemOpen
  \bibfield  {author} {\bibinfo {author} {\bibfnamefont {H.}~\bibnamefont
  {Bombin}}, \bibinfo {author} {\bibfnamefont {G.}~\bibnamefont
  {Duclos-Cianci}}, \ and\ \bibinfo {author} {\bibfnamefont {D.}~\bibnamefont
  {Poulin}},\ }\href@noop {} {\bibfield  {journal} {\bibinfo  {journal}
  {arXiv:1103.4606}} (\bibinfo {year} {2011})}\BibitemShut {NoStop}%
\bibitem [{\citenamefont {Bombin}(2011)}]{Bombin11+}%
  \BibitemOpen
  \bibfield  {author} {\bibinfo {author} {\bibfnamefont {H.}~\bibnamefont
  {Bombin}},\ }\href@noop {} {\bibfield  {journal} {\bibinfo  {journal}
  {arXiv:1107.2707}} (\bibinfo {year} {2011})}\BibitemShut {NoStop}%
\bibitem [{\citenamefont {Bravyi}\ \emph {et~al.}(2006)\citenamefont {Bravyi},
  \citenamefont {Hastings},\ and\ \citenamefont {Verstraete}}]{BHV06}%
  \BibitemOpen
  \bibfield  {author} {\bibinfo {author} {\bibfnamefont {S.}~\bibnamefont
  {Bravyi}}, \bibinfo {author} {\bibfnamefont {M.~B.}\ \bibnamefont
  {Hastings}}, \ and\ \bibinfo {author} {\bibfnamefont {F.}~\bibnamefont
  {Verstraete}},\ }\href@noop {} {\bibfield  {journal} {\bibinfo  {journal}
  {Phys. Rev. Lett.},\ }\textbf {\bibinfo {volume} {97}},\ \bibinfo {pages}
  {050401} (\bibinfo {year} {2006})}\BibitemShut {NoStop}%
\bibitem [{\citenamefont {Bravyi}\ and\ \citenamefont {Terhal}(2009)}]{BT08}%
  \BibitemOpen
  \bibfield  {author} {\bibinfo {author} {\bibfnamefont {S.}~\bibnamefont
  {Bravyi}}\ and\ \bibinfo {author} {\bibfnamefont {B.~M.}\ \bibnamefont
  {Terhal}},\ }\href@noop {} {\bibfield  {journal} {\bibinfo  {journal} {New.
  J. Phys.},\ }\textbf {\bibinfo {volume} {11}},\ \bibinfo {pages} {043029}
  (\bibinfo {year} {2009})}\BibitemShut {NoStop}%
\bibitem [{\citenamefont {Bravyi}\ \emph {et~al.}(2010)\citenamefont {Bravyi},
  \citenamefont {Poulin},\ and\ \citenamefont {Terhal}}]{BPT10}%
  \BibitemOpen
  \bibfield  {author} {\bibinfo {author} {\bibfnamefont {S.}~\bibnamefont
  {Bravyi}}, \bibinfo {author} {\bibfnamefont {D.}~\bibnamefont {Poulin}}, \
  and\ \bibinfo {author} {\bibfnamefont {B.}~\bibnamefont {Terhal}},\
  }\href@noop {} {\bibfield  {journal} {\bibinfo  {journal} {Phys. Rev.
  Lett.},\ }\textbf {\bibinfo {volume} {104}} (\bibinfo {year}
  {2010})}\BibitemShut {NoStop}%
\bibitem [{\citenamefont {Haah}\ and\ \citenamefont {Preskill}(2010)}]{HP10}%
  \BibitemOpen
  \bibfield  {author} {\bibinfo {author} {\bibfnamefont {J.}~\bibnamefont
  {Haah}}\ and\ \bibinfo {author} {\bibfnamefont {J.}~\bibnamefont
  {Preskill}},\ }\href@noop {} {\bibfield  {journal} {\bibinfo  {journal}
  {arXiv:1011.3529}} (\bibinfo {year} {2010})}\BibitemShut {NoStop}%
\end{thebibliography}

%

\end{document}